# Consideration of non-phase-matched nonlinear effects in the design of quasi-phase-matching crystals for optical parametric oscillators

ZIHUA ZHENG[1], ZIWEN TANG[1,2], ZHIYI WEI[3,4,5], JINGHUA SUN[1,5,6]

[1]*Institute of Ultrafast Optics and Photonics, Dongguan University of Technology. Dongguan 523808, China*
[2]*Guangdong Provincial Key Laboratory of Nanophotonic Functional Materials and Devices, South China Normal University, Guangzhou 510006, China*
[3]*Beijing National Laboratory for Condensed Matter Physics, Institute of Physics, Chinese Academy of Sciences, Beijing 100190, China*
[4]*Songshan Lake Materials Laboratory, Dongguan 523808, China*
[5]*Quantum Science Center of Guangdong-HongKong-Macao Greater Bay Area, Shenzhen-Hong Kong International Science and Technology Park, NO.3 Binglang Road, Futian District, Shenzhen, Guangdong*
[6]*sunjh@dgut.edu.cn*

**Abstract:** Femtosecond optical parametric oscillators (OPOs) are widely used in ultrafast nonlinear frequency conversion and quantum information. However, conventional OPOs based on quasi-phase-matching (QPM) crystals have many parasitic non-phase-matched processes which decreases the conversion efficiency. Here, we propose nine-wave coupled equations (NWCEs) to simulate all phase-matched and non-phase-matched interactions in QPM crystals to improve conventional three-wave coupled equations (TWCEs), especially for the situation of high intensity ultrashort pulses and complexly structured crystals. We discuss how to design the poling period of QPM crystal to maximize the conversion efficiency of signal light for a given OPO system. The simulation reveals that the OPO based on chirped periodically poled lithium niobate (CPPLN) with certain chirp rate has higher signal wave conversion efficiency than that of a PPLN, and demonstrates that NWCEs illustrates more details of the pulse evolution in the OPO cavity. Starting from a CPPLN, an aperiodically poled lithium niobate (APPLN) design is available by modifying the domain lengths of the crystal and optimizing the OPO output power via dynamical optimization algorithm. The results show that by using a properly designed APPLN crystal, a 1600nm OPO, when pumped by a femtosecond laser with 1030nm central wavelength, 150 femtosecond pulse duration and 5 GW/cm$^2$ power intensity at the focus, can achieve very efficient output with a signal light conversion efficiency of 50.6%, which is higher than that of PPLN (25.2%) and CPPLN (40.2%). The scheme in this paper will provide a reference for the design of nonlinear QPM crystals of OPOs and will help to understand the complex nonlinear dynamical behavior in OPO cavities.

## 1.Introduction

Laser technology has been developing rapidly in the fields of light-matter interaction, precision measurement and sensing, quantum information, biophotonics [1]-[8], *etc*. However, the wavelength of a laser is limited by the energy level structure of the gain medium. Nonlinear optical frequency conversion techniques were usually adopted to expand laser spectra and even prepare quantum states for wider applications.

Quasi-phase-matching is a widely used nonlinear frequency conversion technique[9]. The most common approach is to apply electric field to periodically invert the domain of a piece of ferroelectric crystal to obtain a periodical poling crystal (PPLN for example). Subsequently, it is found that periodically poled crystal is not the optimal QPM crystal. In 1997, M. M. Fejer *et al.* [10] theoretically proposed the use of CPPLN (shown in Fig. 1). By applying CPPLN, pulse compression can be realized in second harmonic generation (SHG) process [11]. The pulse compression effect in OPO was first reported by T. Beddard *et al.* [12], where a signal pulse with a central wavelength of 3 μm and a pulse width of only 53 fs containing 5 optical cycles was achieved by use of a CPPLN. D. Artigas *et al.*[13] suggested that a CPPLN could be used as a nonlinear crystal for OPO in order to broaden the signal light, thus lowering the pumping threshold as well as suppressing the walk-off effect. Z. Zhang *et al.*[14] investigated the relationship between the spectral width of the OPO signal light, pumped by Fourier transform limited pulses, and the chirp rate of the CPPLN, and by reasonably optimizing the chirp rate of the CPPLN, obtained a broadband signal spectrum spanning from 1350 nm to 1600 nm.

In addition to CPPLN, OPO's nonlinear crystal can also be APPLN with complete freedom of domain design. In 1997, Zhu *et al.* proposed quasi periodically poled lithium niobate (QPPLN) to achieve multiwavelength SHG [15] as well as coupled third harmonic generation (CTHG) [16]. However, the poling of QPPLN is not yet completely freely designed to show the potentials of QPM. In 2000, Gu *et al.* proposed for the first time an APPLN crystal, which breaks through all the limitations of the poling pattern, and the length of each domain can be freely modulated. Efficient QPM of multiple processes can be realized in a single APPLN crystal [17]. Subsequent researches realized multiwavelength parametric amplification [18], CTHG [19], difference frequency generation(DFG) [20], simultaneous parametric down conversion [21], simultaneous matching of multiple processes[22-23], and supercontinuum generation [24-26] by using APPLN crystals. However, the principle of most APPLN designs is to maximize a specific reciprocal lattice vector by modulation of the crystal domain lengths to achieve multi-process phase matching. Those crystals were designed for single pass (a few APPLNs designed for OPOs were also maximized for single-pass gain) and did not take into account the dynamical evolution of the optical pulses as they oscillating in the cavity, nor the effect of non-phase-matched processes. Those may cause notable deviation between theoretical simulation and experimental results.

In this paper, we discuss how to design the poling pattern of a QPM crystal to maximize the efficiency in an OPO cavity. We propose nine-wave coupled equations to simulate all phase-matched and non-phase-matched interactions in QPM crystals to improve conventional three-wave coupled equations for the situations where a complex structure crystal was used in an OPO cavity. NWCEs simulation shows the relationship between chirp rate of the CPPLN and output efficiency of the signal pulse. Then, the NWCEs is also used to design the APPLN crystal to achieve higher efficiency of a 1600nm OPO, when pumped at 1030nm. The results show that the signal light conversion efficiency reaches to 50.6%, which is higher than that of PPLN (25.2%) and CPPLN (40.2%). We believe NWCEs will help us to simulate the $\chi^{(2)}$ nonlinear effects in more details and better accuracy.

## 2.CPPLN and OPO

Conventional QPM crystal PPLN are periodically poled, each crystal domain is of equal length. However, such crystal can only provide a particular reciprocal lattice vector to realize a nonlinear frequency conversion in a narrow spectrum bandwidth. Unlike PPLN, the poling period of CPPLN crystal is changing, and the domain length of the crystal can be either increased or decreased in the beam direction, which are referred to as negatively chirped and positively chirped crystal, respectively, as shown in Figs. 1(a) and (b). The local QPM period, $\Lambda(z)$, is then given by $\Lambda(z) = \dfrac{\Lambda_0}{1+\Lambda_0 D_g z/\pi}$, where $\Lambda_0 = 2\pi/\Delta k$, and $D_g$ referes to *chirp rate* [10].

The Fourier transform limit pulse will generate positively chirped pulse when it passes through the positively chirped crystal and negatively chirped pulse through the negatively chirped crystal.

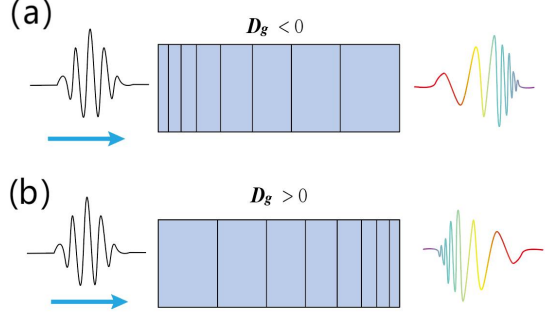

Fig. 1 Schematic diagram of CPPLN crystal. (a) and (b) are negatively chirped crystal and positively chirped crystal, respectively. The Fourier transform limit pulse will generate positively chirped pulse when it passes through the positively chirped crystal and negatively chirped pulse through the negatively chirped crystal.

In this paper, we assume that the OPO (as shown in Fig.2) is pumped by a Yb-doped femtosecond laser with 1030 nm central wavelength, 1.5W average power, 100MHz repetition frequency, and 25 μm focusing spot radius. For simplicity, the output of the laser is assumed as Fourier transform limited Gaussian pulse with full width at half maximum of 150 fs, so the peak power is 5GW/cm$^2$. The time domain electric field of the pulse can be expressed as

$$E(t) = E_0 \exp[-1.385(t/t_0)^2].$$

The OPO is singly resonate for single light and the output coupler is 10%. The central wavelength of the signal light is 1600 nm, initialized from quantum noise.

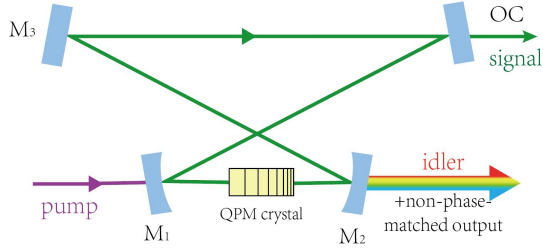

Fig. 2 Schematic diagram of the OPO, where M1 and M2 are curved dichromatic mirrors with focusing length as 50mm, M3 is a plane mirror for 1600nm, and OC is a output coupler with transmission as 10%.

### 3. Nine-wave coupled equations

Due to the high conversion efficiency of an OPO based on a QPM crystal, normally the output lights are very rich in colors, containing phase-matched and non-phase-matched $\chi^{(2)}$ processes. In this scenario, the conventional TWCEs [27] are not sufficient to resolve the dynamics of the optical fields, hence, we propose NWCEs to describe comprehensively the nonlinear interactions in the crystal. The main interaction is

$$\omega_1 \to \omega_2 + \omega_3,$$

where $\omega_1$, $\omega_2$, $\omega_3$ refer to the frequency of pump, signal and idler, respectively. The secondly interactions occur between the three optical fields of the main reaction, including SHG

$$\omega_1 + \omega_1 \to \omega_4,$$
$$\omega_2 + \omega_2 \to \omega_5,$$
$$\omega_3 + \omega_3 \to \omega_6,$$

sum-frequency generation (SFG)
$$\omega_1 + \omega_2 \to \omega_7,$$
$$\omega_1 + \omega_3 \to \omega_8,$$

and difference-frequency generation (DFG)
$$\omega_2 \to \omega_3 + \omega_9.$$

Of course, one may come up with even further interactions between these nine waves, but they are too weak to be considered in real cases. Let $E_j$ refer to the optical field of $\omega_j$ (j=1~9, the same below), $n_j$ the refractive index of the crystal, and $k_j = n_j \omega_j / c$ the wave vectors. $\Delta k_{abc} = k_a - k_b - k_c$ refers to the phase mismatch between the three wave vectors $k_a$, $k_b$, and $k_c$. For example, $\Delta k_{411} = k_4 - k_1 - k_1$, $\Delta k_{239} = k_2 - k_3 - k_9$, and so on. Let the z-axis be the beam direction. The nonlinear coefficient of the crystal in this direction is $d_{eff}$. The second-order dispersion effects is considered. The conventional TWCEs describe the evolution of the three optical fields $E_1$, $E_2$, and $E_3$ of the main interactions as

$$\frac{\partial E_1}{\partial z} + \left(\frac{1}{v_{g1}}\frac{\partial}{\partial t} + \frac{i\beta_1}{2}\frac{\partial^2}{\partial t^2}\right)E_1 = C_1 E_2 E_3 e^{-i\Delta k_{123} z},$$

$$\frac{\partial E_2}{\partial z} + \left(\frac{1}{v_{g2}}\frac{\partial}{\partial t} + \frac{i\beta_2}{2}\frac{\partial^2}{\partial t^2}\right)E_2 = C_2 E_1 E_3^* e^{i\Delta k_{123} z},$$

$$\frac{\partial E_3}{\partial z} + \left(\frac{1}{v_{g3}}\frac{\partial}{\partial t} + \frac{i\beta_3}{2}\frac{\partial^2}{\partial t^2}\right)E_3 = C_3 E_1 E_2^* e^{i\Delta k_{123} z}.$$

where $C_j = \frac{i\omega_j d(z)}{cn_j}$ are coupling coefficients, $v_{gj}$ and $\beta_j$ are the group velocity and second-order dispersion of the optical field $E_j$, respectively.

The intracavity peak power intensity of a femtosecond OPO reaches typically above 10 GW/cm², so that the secondary effects, such as SHG, SFG and DFG, are nonnegligible. If the traditional TWCEs are used for simulation, the signal light conversion efficiency obtained will be higher than the actual value. In addition, for aperiodically poled nonlinear crystals, which provide multiple reciprocal lattice vectors for multiple frequency conversion processes, there are strong correlations between different frequency conversion processes. The TWCEs cannot reveal the rich dynamical evolution of the optical fields in the cavity.

The evolution of the nine optical fields satisfies the following nine-wave coupled equations:

$$\frac{\partial E_1}{\partial z} + \left(\frac{1}{v_{g1}}\frac{\partial}{\partial t} + \frac{i\beta_1}{2}\frac{\partial^2}{\partial t^2}\right)E_1 = C_1(E_2 E_3 e^{-i\Delta k_{123}z} + E_1^* E_4 e^{i\Delta k_{411}z} + E_2^* E_7 e^{i\Delta k_{712}z} + E_3^* E_8 e^{i\Delta k_{813}z})$$

$$\frac{\partial E_2}{\partial z} + \left(\frac{1}{v_{g2}}\frac{\partial}{\partial t} + \frac{i\beta_2}{2}\frac{\partial^2}{\partial t^2}\right)E_2 = C_2(E_1 E_3^* e^{i\Delta k_{123}z} + E_2^* E_5 e^{i\Delta k_{522}z} + E_1^* E_7 e^{i\Delta k_{712}z} + E_3 E_9 e^{-i\Delta k_{239}z})$$

$$\frac{\partial E_3}{\partial z} + \left(\frac{1}{v_{g3}}\frac{\partial}{\partial t} + \frac{i\beta_3}{2}\frac{\partial^2}{\partial t^2}\right)E_3 = C_3(E_1 E_2^* e^{i\Delta k_{123}z} + E_3^* E_6 e^{i\Delta k_{633}z} + E_1^* E_8 e^{i\Delta k_{813}z} + E_2^* E_9 e^{i\Delta k_{239}z})$$

$$\frac{\partial E_4}{\partial z} + \left(\frac{1}{v_{g4}}\frac{\partial}{\partial t} + \frac{i\beta_4}{2}\frac{\partial^2}{\partial t^2}\right)E_4 = \frac{1}{2}C_4 E_1^2 e^{-i\Delta k_{411}z}$$

$$\frac{\partial E_5}{\partial z} + \left(\frac{1}{v_{g5}}\frac{\partial}{\partial t} + \frac{i\beta_5}{2}\frac{\partial^2}{\partial t^2}\right)E_5 = \frac{1}{2}C_5 E_2^2 e^{-i\Delta k_{522}z}$$

$$\frac{\partial E_6}{\partial z} + \left(\frac{1}{v_{g6}}\frac{\partial}{\partial t} + \frac{i\beta_6}{2}\frac{\partial^2}{\partial t^2}\right)E_6 = \frac{1}{2}C_6 E_3^2 e^{-i\Delta k_{633}z}$$

$$\frac{\partial E_7}{\partial z} + \left(\frac{1}{v_{g7}}\frac{\partial}{\partial t} + \frac{i\beta_7}{2}\frac{\partial^2}{\partial t^2}\right)E_7 = C_7 E_1 E_2 e^{-i\Delta k_{712}z}$$

$$\frac{\partial E_8}{\partial z} + \left(\frac{1}{v_{g8}}\frac{\partial}{\partial t} + \frac{i\beta_8}{2}\frac{\partial^2}{\partial t^2}\right)E_8 = C_8 E_1 E_3 e^{-i\Delta k_{813}z}$$

$$\frac{\partial E_9}{\partial z} + \left(\frac{1}{v_{g9}}\frac{\partial}{\partial t} + \frac{i\beta_9}{2}\frac{\partial^2}{\partial t^2}\right)E_9 = C_9 E_2 E_3^* e^{i\Delta k_{239}z}$$

We use the split-step Fourier method [28] to solve above equations.

In order to verify the correctness of the nine-wave coupled equations, we choose a PPLN crystal first in the OPO cavity described above. The length of the PPLN is 0.5 mm with a poling period of 30.45 μm matching for the process of parametric down conversion from 1030 nm to 1600 nm. The conversion generates not only signal of 1600nm and the idler of 2891 nm, but also SHGs (515nm, 800nm, 1446nm), SFGs (627nm, 759nm), and DFG (3583 nm), as shown in Fig. 3.

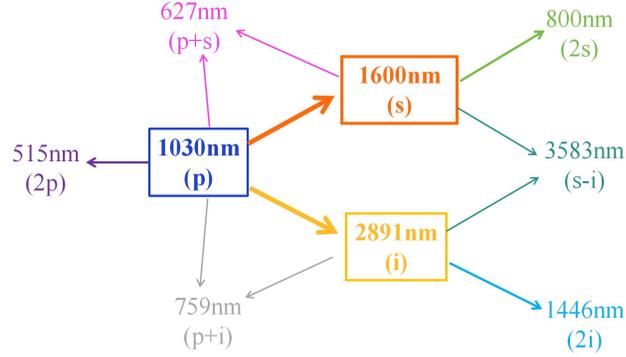

Fig. 3  Schematic diagram of the nine-wave interaction relationship.

We calculate the output signal optical field and conversion efficiency using NWCEs and TWCEs, respectively, and the results are shown in Fig. 4, where (a) and (b) are the outputs of the OPO calculated using TWCEs and NWCEs, respectively; (c) and (d) are the evolution curves of the nine-waves versus the signal round trips.

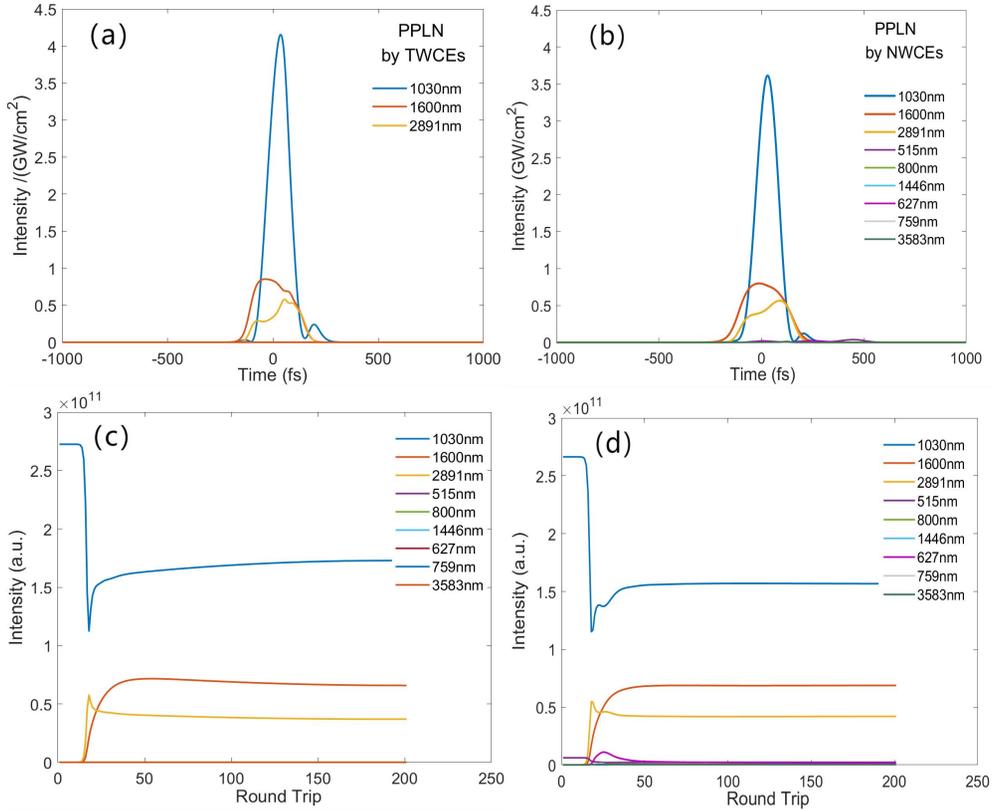

Fig.4 (a) and (b) are the PPLN OPO outputs calculated by using TWCEs and NWCEs respectively; (c) and (d) are the evolution curves of all lights simulated by using TWCEs and NWCEs, respectively.

The simulated conversion efficiencies are 23.9% and 25.2% from a PPLN OPO by using TWCEs and NWCEs, respectively. The conversion efficiencies and pulse envelopes obtained from these two sets of equations are very similar, indicating the correctness of the nine-wave coupled equations.

When a CPPLN with a chirp rate of $D_g=4\times 10^{-5}\mu m^{-2}$ is used in the OPO, with other conditions unchanged, it can be seen that the simulation results diverge obviously, as shown in Fig. 5. Due to the signal light oscillating in the cavity is very strong (up to 15 GW/cm$^2$), it nonlinearly interacts with the pump, generating intensive sum-frequency wave (627nm), as shown in Fig. 5(b), which is neglected in the conventional TWCEs simulation (Fig. 5(a)). As a result, the simulation of the TWCE overestimates the conversion efficiency of the signal light to 56.6% (see Fig. 5(c)), in contrast, it is only 33.8% from NWCEs (see Fig. 5(d)).

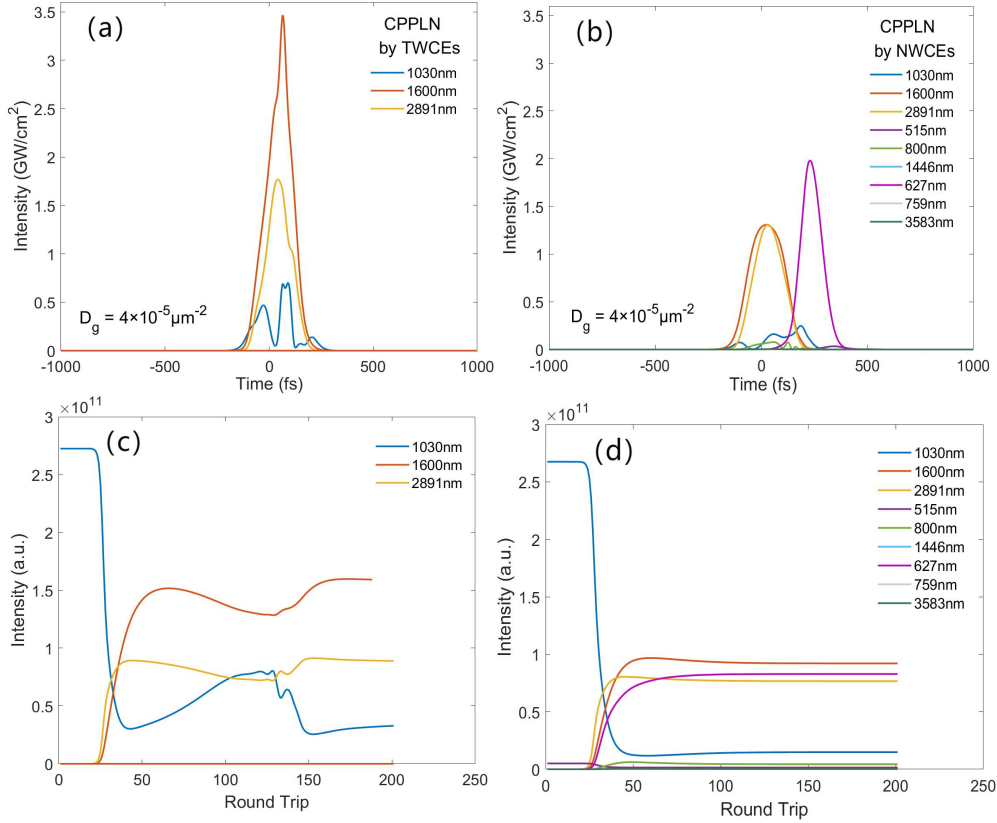

Fig.5 (a) and (b) are OPO outputs calculated by using TWCEs and NWCEs, respectively when the nonlinear crystal is a CPPLN with a chirp rate of $4\times10^{-5}\mu m^{-2}$; (c) and (d) are the evolution curves of all lights simulated by using TWCEs and NWCEs, espectively.

As shown in Fig. 5(c) and (d), if only the three-wave interaction is considered, the system is very unstable, in contrast, using the nine-wave coupled equations, the system gradually evolves to a stable state, which is consistent with the experimental situation.

## 4. Optimization of CPPLN

With the establishment of NWCEs, we can simulate pulse interactions in an OPO cavity based on a QPM crystal accurately, especially in the situation of wide bandwidth and high intensity. This helps us investigate the conversion efficiency of an OPO based on CPPLN with different chirp rate $D_g$.

We assume that the length of the nonlinear crystal in the OPO is fixed at 0.5 mm. The length of the central domain in the CPPLN is the coherence length of the central wavelength of the main interaction, which is 15.2μm for 1030nm to 1600nm down conversion. Fig. 6 shows the signal conversion efficiency of the OPO when the CPPLN has different chirp rates.

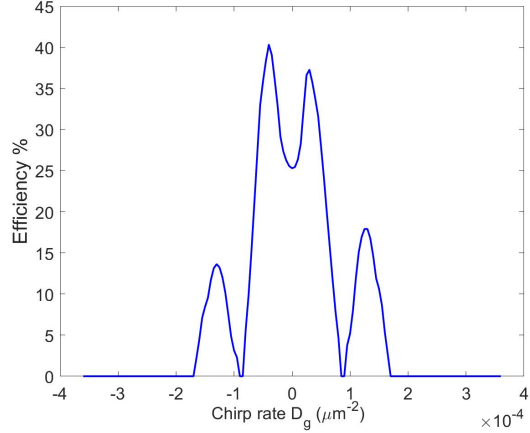

Fig. 6 OPO signal light conversion efficiency versus CPPLN chirp rate.

The curve in Fig. 6 shows some kind of symmetry around $D_g$=0. That means CPPLNs with small chirp rate, either negative or positive, have higher conversion efficiencies than a PPLN. The maximum conversion efficiency reaches to 40.2%. when $D_g$=-4×10$^{-5}$μm$^{-2}$. For crystals with equal absolute values of chirp rate, the signal light conversion efficiencies are very close. This may be due to the fact that the two crystals with equal $|D_g|$ just have different beam direction, but with the same poling periods, and thus the same reciprocal lattice vector.

However, the conversion efficiency curve is not exactly symetric, it is worth researching the performance of the OPO in details when the crystal has negative and positive chirp. Taking $D_g$=-4×10$^{-5}$μm$^{-2}$ and $D_g$=4×10$^{-5}$μm$^{-2}$ as examples, their output pulses, simulated by applying NWCEs, are shown in Fig. 7 (a) and (b), respectively. It can be seen that the 627nm sum-frequency light produced when $D_g$=4×10$^{-5}$μm$^{-2}$ is very strong. This is because the CPPLN crystal for 1030nm→1600nm+2891nm interaction has a reciprocal lattice vector of the third-order QPM as 0.619μm$^{-1}$, which is very close to the phase mismatch of 1030nm+1600nm→627nm process (0.542μm$^{-1}$), as shown in Fig. 7(c), and hence 627nm SFG light obtains quite high gain along such a CPPLN crystal. But it is also obvious that the 627nm generated from the $D_g$=-4×10$^{-5}$μm$^{-2}$ crystal is weaker than that of $D_g$=4×10$^{-5}$μm$^{-2}$. This may be explained as that the domain length of the negatively chirped CPPLN is gradually decreasing, with longer domains at the beginning. Shorter domain length is more favorable for sum-frequency generation of 627 nm, along with higher pumping power at the beginning, generating higher 627nm power than the case of a positively chirped CPPLN.

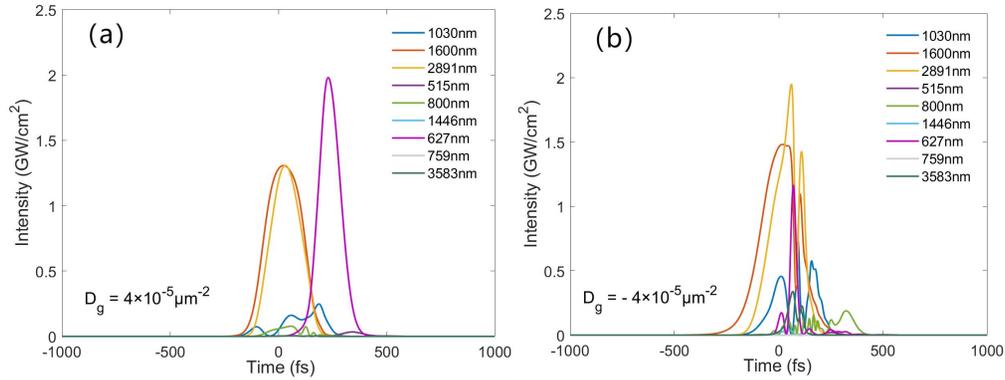

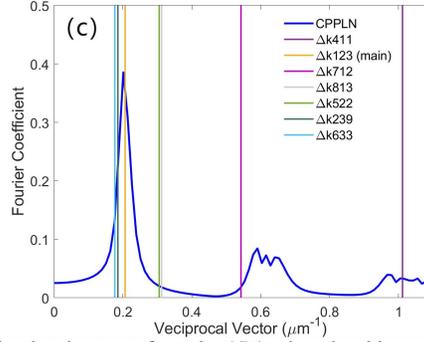

Fig.7. (a) and (b) are simulated outputs from the OPO when the chirp rate of the CPPLN are $D_g=4\times10^{-5}\mu m^{-2}$ and $-4\times10^{-5}\mu m^{-2}$, respectively. (c) is the distribution of the reciprocal lattice vectors provided by the CPPLN with $|D_g|=-4\times10^{-5}\mu m^{-2}$, along with the reciprocal lattice vectors of the main interaction ($\Delta k_{123}$) and secondary interactions shown as vertical lines.

## 5. Optimization of APPLN

As discussed in the above section, the CPPLN crystal can enhance the nonlinear conversion efficiency of the OPO significantly, comparing with a PPLN OPO. But it is also obvious that there are many unwanted secondary outputs, wasting the pumping energy and disturbing the pulse interactions.

NWCEs bring about the possibility of breaking the limit of linear change of the domain lengths, and designing a QPM crystal with fully aperiodically poled QPM crystal for higher conversion efficiency by suppressing secondary interactions or for multiple nonlinear processes.

APPLNs were adopted for pulse shaping[10,29,30], multiprocessing frequency conversion[22], supercontinuum generation[24-26], and other fields. Here we optimize the design of APPLN crystal to enhance the efficiency of OPO by using NWCEs.

In our simulation, the initialization of the APPLN starts from the optimal CPPLN which has $D_g=4\times10^{-5}\mu m^{-2}$ and $L=0.5$ mm. The OPO reaches to a steady state after 100-200 round trips, when the intensity of the signal light is $I_s$.

During the training process, one crystal domain of the APPLN is randomly changed each time by increasing or decreasing its length of 2% to re-simulate the OPO operation until a new stable state is reached and get a new signal intensity $I_s'$. If $I_s'>I_s$, this new set of crystal domain distribution is accepted, otherwise rejected. For a given number of repeating of the above process, the program will output the optimal crystal domain in the end.

However, in this scheme, it takes 100-200 turns of the new OPO system to reach a steady state for every domain modification, which is very time-consuming when a great number of times of the domain modification are required. In order to reduce the computing time, we propose a new method, *dynamical optimization algorithm*, in which we only run a small number of turns (such as 10 turns) based on previous OPO computation results, instead of restarting the OPO simulation from quantum noise after every modification of the crystal domain. This is reasonable since only one of the domains is changed by 2% in length each time, the new APPLN is very close to the previous one, so that its output should be stable enough after a small number of turns.

The dynamical optimization algorithm is summarized below:
(1) Initialize OPO program, run the OPO simulation 200 round trips.
(2) For each search (modifying the length of one random domain): run 10 round trips, record $I_s$ of the signal light; change a crystal domain, run 10 turns, record $I_s'$ of the signal light, and accept the new crystal domain if $I_s' > I_s$.
(3) A total of 1000 searches.

Running the above algorithm gives an APPLN optimization result as shown in Fig.8.

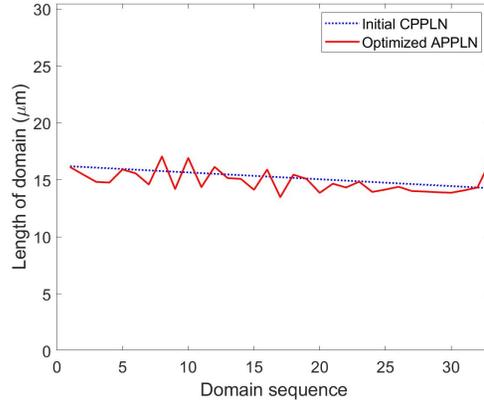

Fig. 8 Comparison of the optimized domain distribution obtained after 1000 searches (solid red) with the initial CPPLN domain (dashed blue).

Once the poling distribution of the APPLN is optimized, it is put into the OPO simulation program to analyze the signal outputs. The temporal and spectral characteristics of output optical fields are shown in Fig. 9(a1) and (b1), respectively. As a comparison, the outputs of the OPO with the CPPLN carrying chirp rate of $D_g=4\times10^{-5}\mu m^{-2}$ is also simulated and presented in Fig. 9 (a2) and (b2). Fig. 9 (c1) and (c2) show the pulses evolving in the OPO cavities. It is evident that the signal pulse is notably higher, and the SFG 627nm is significantly lower from the APPLN OPO comparing to the CPPLN OPO.

Table 1 lists a detailed comparison of the conversion efficiencies of all the nine waves between the APPLN OPO and the CPPLN OPO. The former's performance is better than the latter in terms of enhancing phase-matched conversion and compressing non-phase-matched ones.

Table 1 Conversion efficiencies of the nine pulses when the OPO of CPPLN and APPLN are stabilized

|  | 1030nm | 1600nm | 2891nm | 515nm | 800nm | 1446nm | 627nm | 759nm | 3583nm |
|---|---|---|---|---|---|---|---|---|---|
| APPLN | 8.8% | 50.6% | 34.0% | 0.1% | 1.9% | 0.3% | 3.1% | 0.0% | 1.2% |
| CPPLN | 5.5% | 33.8% | 28.1% | 0.5% | 1.6% | 0.0% | 30.4% | 0.0% | 0.1% |

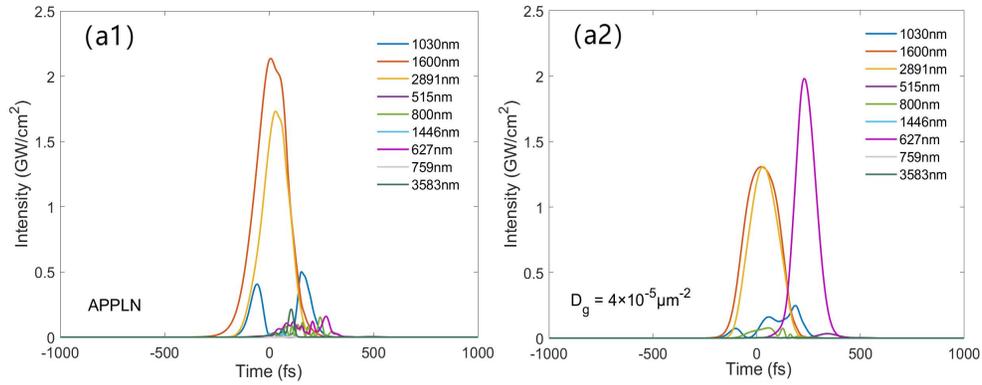

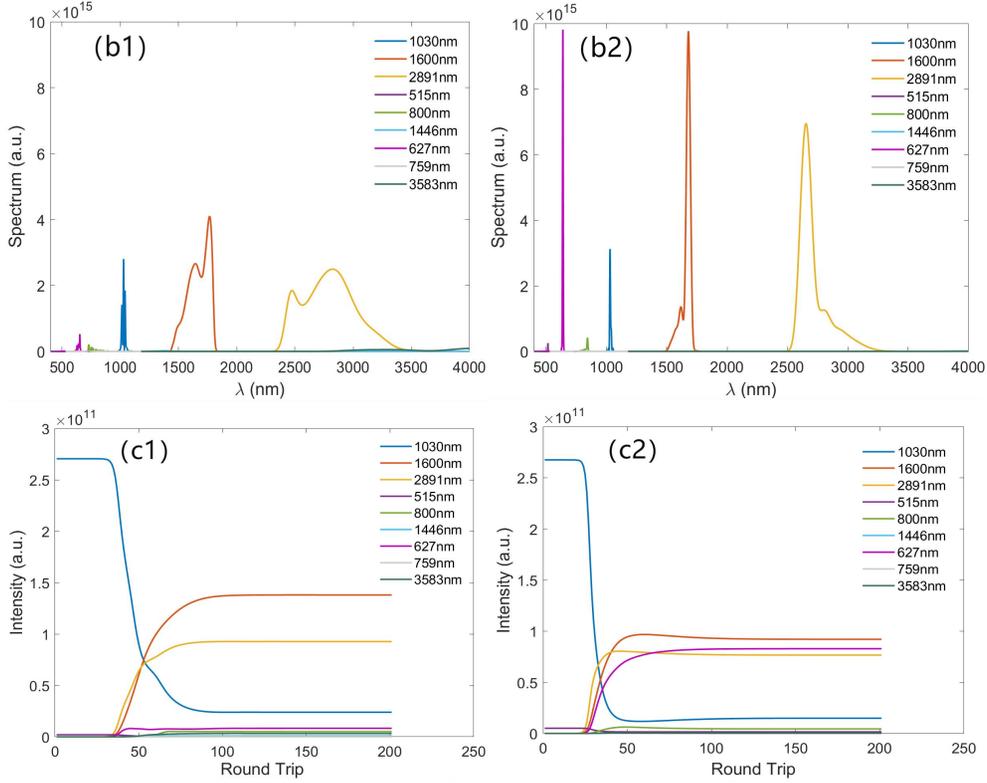

Fig. 9 The outputs in time domain (a1, a2) and spectrum domain (b1, b2), and the pulse intensity evolution (c1, c2) from the OPOs with the APPLN and the CPPLN ($D_g=4\times10^{-5}\mu m^{-2}$), respectively.

In order to further understand the principle of the suppression of non-phase-matched waves by aperiodically poling the crystal domains, we calculate the reciprocal lattice vector distribution of the APPLN and compare it with the CPPLN ($D_g=4\times10^{-5}\mu m^{-2}$), as shown in Fig.10. It can be seen that the reciprocal lattice vector curve of the APPLN has an obvious enhancement for the main interactions ($\omega_1+\omega_2\rightarrow\omega_3$, corresponding to $\Delta k_{123}$) and avoids the non-phase-matched interactions of $\omega_3+\omega_3\rightarrow\omega_6$, $\omega_2\rightarrow\omega_3+\omega_9$, and $\omega_1+\omega_2\rightarrow\omega_7$ (corresponding to $\Delta k_{633}$, $\Delta k_{239}$ and $\Delta k_{712}$, respectively).

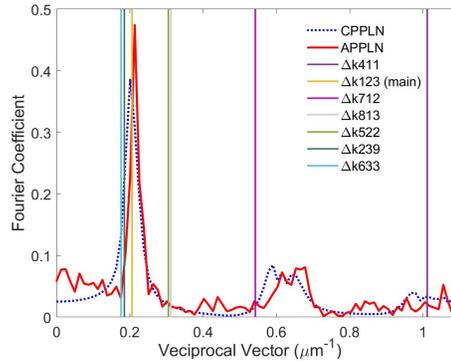

Fig.10 Distribution of reciprocal lattice vectors provided by the CPPLN with $D_g=4\times10^{-5}\mu m^{-2}$ (blue) versus the optimal APPLN (red) crystal. The vertical lines are the reciprocal lattice vectors of the main reaction ($\Delta k_{123}$) and the series of secondary conversions.

The APPLN crystal optimized previously provides higher conversion efficiency and wider spectrum bandwidth for signal light (see Fig. 9), but not necessarily generates shorter signal pulse since it might be chirped. It is worth investigating the Fourier transform limit of the signal pulse width from the OPO. In general, people use prism pairs [32-33] or grating pairs [34-35] to compensate the positive linear chirp of ultrafast optical pulses to make the pulse width as close as possible to Fourier transform limit, especially the wavelength is below 1.5mm region. For this purpose, we simulate the pulse compression of the signal light by intoducing group delay dispersion (GDD).

The envelope of the pulse after passing through a prism pair can be expressed as

$$E'(t) = F^{-1}\{F[E(t)] \cdot \exp[i \cdot GDD \cdot (\omega - \omega_0)^2/2]\},$$

where $E(t)$ is the signal pulse envelope before entering the prism pair, $\omega_0$ is the central frequency of the pulse, and $F$, $F^{-1}$ are the Fourier transform and inverse Fourier transform operators, respectively. By changing the GDD value to the output signal pulse, its duration reaches to a minimum when GDD=-1000fs$^2$, as shown in Fig.11 (a). At this point, the pulse duration is compressed to 25 fs (Fig. 11(b)), which is much shorter than the uncompressed width of 173fs.

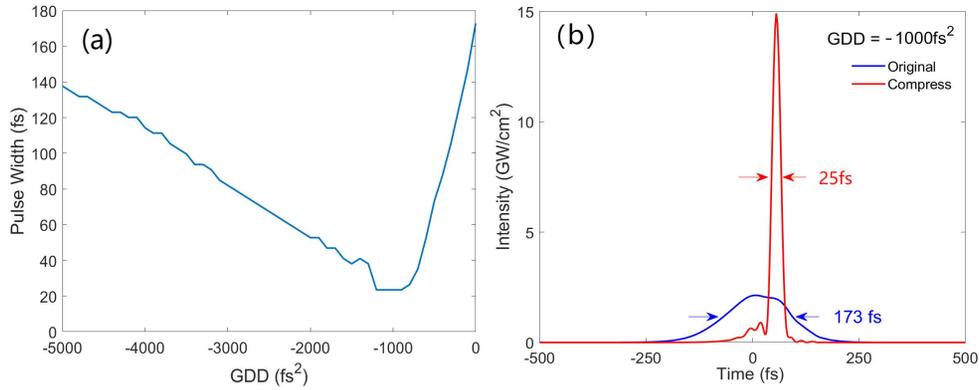

Fig.11 Pulse compression of the signal light from the APPLN OPO. (a) Peak power intensity versus GDD; (b) compressed pulse (red) when the GDD is -1000 fs$^2$ compared with the uncompressed pulse (blue).

## 6. Conclusion

We have discussed how to design the poling period of a QPM crystal to maximize the conversion efficiency of the signal light for a given OPO system. We propose nine-wave coupled equations to analyze comprehensively the evolution of the optical fields interacting in the QPM crystal. With the NWCEs, We firstly investigated the performance of CPPLNs with different grating chirp rate $D_g$, and found that a CPPLN with certain $D_g$ value has higher signal light conversion efficiency than a PPLN, and also discovered that CPPLNs with $+|D_g|$ and $-|D_g|$ ($|D_g|=4 \times 10^{-5}\mu m^{-2}$ in our simulation) have very different performance in terms of secondary interactions (non-phase-matched interactions). Then, we designed an APPLN crystal by optimizing the domains originating from the CPPLN of $D_g=4 \times 10^{-5}\mu m^{-2}$), and compared the outputs from the OPOs based on these two crystals. The conversion efficiency of the APPLN OPO, from 1030nm pump to 1600nm signal, reaches to 50.6%, along with obvious suppression of non-phase-matched interactions, which is higher than that from the CPPLN (40.2%) and the PPLN (25.2%). Further, since the APPLN OPO generates wider signal spectrum than that from the CPPLN OPO, the signal pulse duration can be compressed to 25 fs, in contrast to 45fs from the latter, when the pump pulse width is as long as 150fs.

Conventional QPM crystal design is to maximize a specific effective nonlinear coefficient through optimizing the domain distribution of the crystal, in which the information of the nonlinear dynamics of the optical fields evolving in the oscillator is lost. Reciprocal lattice vector can not completely determine the distribution of the optical,as mentioned in the analysis of CPPLN that the same reciprocal lattice vector may correspond to different crystals and produce very different optical fields. The dynamical optimization algorithm we have presented is to analyse the evolution of the optical fields in the OPO cavity, and to optimize the output signal conversion efficiency dynamically. With the NWCEs, we can obtain more detailed information about the pulses evolving in the cavity, not only the three pulses of the main interactions, but also those of non-phase-matched secondary interactions, and can design an APPLN with high signal conversion efficiency and obvious suppression of unwanted secondary lights. The design and simulation results are more reasonable than those by conventional three-wave coupled equations since the secondary interactions are nonnegligible in an femtosecond OPO where pulse peak intensity can be higher than 10 GW/cm$^2$. In the future studies, we will try to customize an APPLN crystal based on our simulation above to verify the effectiveness experimentally.

For a femtosecond OPO, there are more effects such as group velocity delay and group velocity dispersion of the cavity (not the crystal only) could be considered to analyse the optical outputs in more complex situations, but it also increases the computational efforts. Simulation with a more powerful computer or even a quantum computer[36] will enable more detailed APPLN crystal design for high efficiency and multipurpose OPO with consideration of non-phase-matched interactions. New technologies for QPM crystal manufacturing with very high domain fineness[37] are emerging, making meticulous APPLN design feasible and valuable for applications.

**Funding.** National Key Research and Development Program of China (2022YFB4601103); Guangdong Basic and Applied Basic Research Foundation (2020B1515120041).

**Acknowledgments.** We gratefully acknowledge the helpful discussions with Prof. Usman K. Sapeav from the National University of Uzbekistan and Tashkent State Technical University.

**Disclosures.** The authors declare no conflicts of interest.

**Data availability.** Data underlying the results presented in this paper are not publicly available at this time but may be obtained from the authors upon reasonable request.

See Supplementary Materials for supporting content.

# Supplementary information for:

# Consideration of non-phase-matched nonlinear effects in the design of quasi-phase-matching crystals for optical parametric oscillators

## 1. Definition of CPPLN and APPLN

APPLN and CPPLN have different meanings in different articles. Here we clarify the meanings of these two concepts. For CPPLN, the local QPM period $\Lambda(z)$ is given by $\Lambda(z) = \frac{\Lambda_0}{1 + \Lambda_0 D_g z/\pi}$, where $\Lambda_0 = 2\pi/\Delta k$, and $D_g$ refers to chirp rate. Other non-uniform poling QPM crystals that do not follow this polarization pattern are called APPLN.

In some literature, such as [1], *linearly* chirped polarized crystal is considered as a kind of aperiodic polarized crystal, which is called APPLN, but it actually is not a *freely-designed* APPLN. So we call the crystal a CPPLN with a chirp rate of $D_g$ in this paper.

In some other article, such as [2], the domain length is *nonlinearly* chirped poling. To avoid confusion, we refer to such crystals as APPLN, rather than CPPLN as the original article.

## 2. Verification of our simulation

In order to verify the validity of our simulation codes used in this work, we choose an example where analytical solution is existing, solve it by split-step Fourier method (the method which our simulation is based on) and compare the result with that of the analytical solution.

In general, three-wave coupled equations are nonlinear partial differential equations and almost impossible to get the analytical solution. However, for cw laser, the dispersion can be ignored. In the case of quasi-phase matching, the phase mismatch $\Delta k=0$ and the effective nonlinear coefficient $\chi_{eff} = \frac{2}{\pi}\chi^{(2)}$, the three-wave coupled equations for sum-frequency generation (SFG) $\omega_1+\omega_2=\omega_3$ can be written as

$$\frac{\partial}{\partial z}E_1 = \frac{i\omega_1\chi_{eff}}{2n_1 c}E_2^* E_3,$$

$$\frac{\partial}{\partial z}E_2 = \frac{i\omega_2\chi_{eff}}{2n_2 c}E_1^* E_3,$$

$$\frac{\partial}{\partial z}E_3 = \frac{i\omega_3\chi_{eff}}{2n_3 c}E_1 E_2,$$

where $E_j$ refers to the optical field of $\omega_j$ ($j=1\sim3$ respectively, the same below), $n_j$ is the refractive index of the crystal, $c$ is the speed of light in a vacuum and $i$ is the imaginary unit.

The analytical solution was given in the 1960s by J.A. Arrmstrong and N. Boembergen [3]. Hu *et al*. gave a form that is easier to compute.[4].

Defining energy flow $W = c^2\left[\frac{k_1|E_{10}|^2}{\omega_1} + \frac{k_2|E_{20}|^2}{\omega_2}\right]$, then the analytical solutions of the above equations are

$$|E_1(z)| = \left(\frac{c^2 k_1}{\omega_1^2 W}\right)^{-\frac{1}{2}} u_1(z),$$

$$|E_2(z)| = \left(\frac{c^2 k_2}{\omega_2^2 W}\right)^{-\frac{1}{2}} u_2(z),$$

$$|E_3(z)| = \left(\frac{c^2 k_3}{\omega_3^2 W}\right)^{-\frac{1}{2}} u_3(z),$$

where

$$u_1^2(z) = u_1^2(0) - u_2^2(0) sn^2\left(\frac{z}{z_0}, \gamma\right),$$

$$u_2^2(z) = u_2^2(0) - u_2^2(0) sn^2\left(\frac{z}{z_0}, \gamma\right),$$

$$u_3^2(z) = u_2^2(0) sn^2\left(\frac{z}{z_0}, \gamma\right),$$

with boundary condition

$$u_1(0) = \left|\left(\frac{c^2 k_1}{\omega_1^2 W}\right)^{\frac{1}{2}} E_1(0)\right|,$$

$$u_2(0) = \left|\left(\frac{c^2 k_2}{\omega_2^2 W}\right)^{\frac{1}{2}} E_2(0)\right|.$$

Parameter $\gamma = \dfrac{u_2(0)}{u_1(0)}$, $z_0 = \left[\dfrac{|E_1(0)|\chi_{eff}}{c^2}\left(\dfrac{\omega_2^2 \omega_3^2}{k_2 k_3}\right)^{\frac{1}{2}}\right]^{-1}$, and *sn* is the Jacobi elliptic function.

To test the effectiveness of our codes, the conversion efficiency of SHG is calculated and compared with the analytical solution. It is assumed that the wavelength of the fundamental is 1040 nm and the peak power intensity is 0.1 GW/cm². When the fundamental wave enters the PPLN crystal, it undergoes frequency doubling to 520 nm. The length of the PPLN crystal is 1 mm, and the conversion efficiency curve of the SHG obtained by our codes based on the split-step Fourier method is shown in Figure S1. It can be seen that our numerical solution agrees well with the analytical solution, proving the validity of our codes.

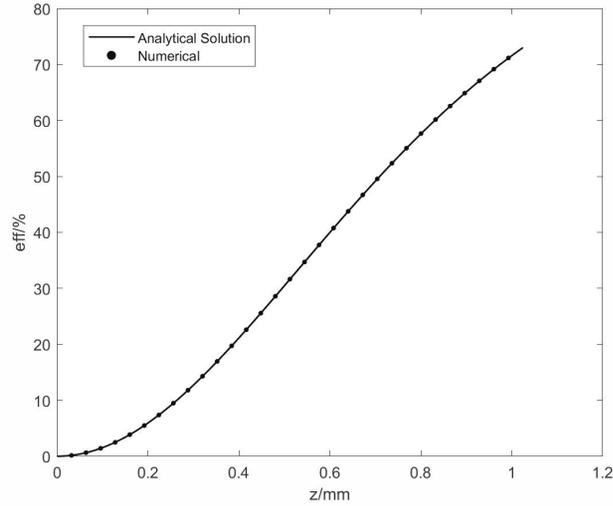

Figure S1 Conversion efficiency of SHG, where the dots are numerical solutions based on the codes used in this paper, and the solid line is the analytical solution of [4].

## 3. Theoretical upper limit of optical conversion efficiency of OPO signal.

If the 1030nm pump light is completely converted into 1600nm signal and 2891nm idler, there is no other loss except output coupling, and no secondary frequency conversions, then the theoretical upper bound (quantum efficiency) of signal light efficiency is 1030/1600 ≈ 64.37%.

## 4. Inverse lattice vector distribution of parasitic processes in OPO with different signal wavelength

The evolution of the 1600nm OPO pumped by 1030nm is simulated by using the nine-wave coupled equations in this paper. Due to the strong signal light oscillating in the cavity, there are some parasitic processes such as SHG and SFG of the pump, signal and idle, etc. The traditional three-wave coupled equations cannot accurately simulate these processes. In fact, since the phase mismatches of the parasitic (secondary) processes are often close to the reciprocal vectors of the odd orders of the QPM of the main interaction, the parasitic processes often appear. Figure S2 shows distribution of the reciprocal vectors of the main interaction and parasitic processes in the OPO pumped by 1030nm (Yb lasers, see Figure S2 (a)) and 800nm (Ti:sapphire laser, see Figure S2 (b)) to different signal wavelengths, respectively.

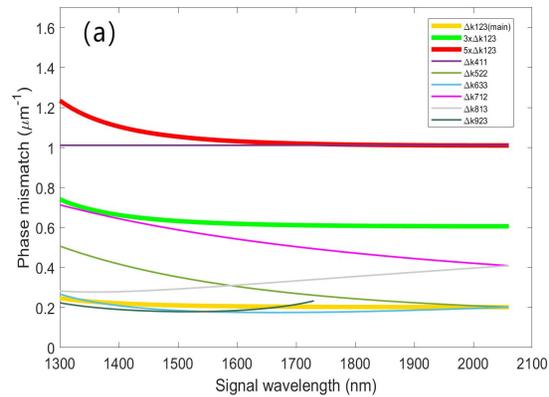

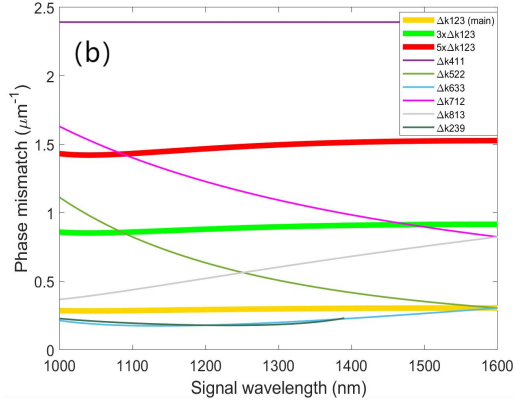

Figure S2 Curves of the phase mismatching of nine waves vs the signal wavelengths of OPOs pumped at 1030 nm (a) and at 800 nm (b), respectively.

The pooling period $\Lambda$ of the QPM crystal is usually around $2\pi/\Delta k_{123}$ in an OPO, the processes with reciprocal vector around $\Delta k_{123}$, $3\Delta k_{123}$, $5\Delta k_{123}$ will be significant enhance, and the nine-wave coupled equations can be used to describe the detailed dynamics.

## 5. Nine-wave coupled equations with χ$^{(3)}$ terms.

Reference [5] presents a three-wave coupled equations with third-order nonlinear terms, which can be rewritten as

$$\frac{\partial E_1}{\partial z} + \left(\frac{1}{v_{g1}}\frac{\partial}{\partial t} + \frac{i\beta_1}{2}\frac{\partial^2}{\partial t^2}\right)E_1 = C_1 E_2 E_3 e^{-i\Delta k_{123} z} + i\frac{3\omega_1 \chi^{(3)}}{8n_1 c} E_1 (S - |E_1|^2),$$

$$\frac{\partial E_2}{\partial z} + \left(\frac{1}{v_{g2}}\frac{\partial}{\partial t} + \frac{i\beta_2}{2}\frac{\partial^2}{\partial t^2}\right)E_2 = C_2 E_1 E_3^* e^{i\Delta k_{123} z} + i\frac{3\omega_2 \chi^{(3)}}{8n_2 c} E_2 (S - |E_2|^2),$$

$$\frac{\partial E_3}{\partial z} + \left(\frac{1}{v_{g3}}\frac{\partial}{\partial t} + \frac{i\beta_3}{2}\frac{\partial^2}{\partial t^2}\right)E_3 = C_3 E_1 E_2^* e^{i\Delta k_{123} z} + i\frac{3\omega_3 \chi^{(3)}}{8n_3 c} E_3 (S - |E_3|^2).$$

Here $S = 2(|E_1|^2 + |E_2|^2 + |E_3|^2)$.

Similarly, we can also add third-order nonlinearities to the nine-wave coupled equations.

$$\frac{\partial E_1}{\partial z} + \left(\frac{1}{v_{g1}}\frac{\partial}{\partial t} + \frac{i\beta_1}{2}\frac{\partial^2}{\partial t^2}\right)E_1 = C_1\left(E_2 E_3 e^{-i\Delta k_{123}z} + E_1^* E_4 e^{i\Delta k_{411}z} + E_2^* E_7 e^{i\Delta k_{712}z} + E_3^* E_8 e^{i\Delta k_{813}z}\right)$$
$$+ i\frac{3\omega_1 \chi^{(3)}}{8n_1 c} E_1\left(S - |E_1|^2\right)$$

$$\frac{\partial E_2}{\partial z} + \left(\frac{1}{v_{g2}}\frac{\partial}{\partial t} + \frac{i\beta_2}{2}\frac{\partial^2}{\partial t^2}\right)E_2 = C_2\left(E_1 E_3^* e^{i\Delta k_{123}z} + E_2^* E_5 e^{i\Delta k_{522}z} + E_1^* E_7 e^{i\Delta k_{712}z} + E_3 E_9 e^{-i\Delta k_{239}z}\right)$$
$$+ i\frac{3\omega_2 \chi^{(3)}}{8n_2 c} E_2\left(S - |E_2|^2\right)$$

$$\frac{\partial E_3}{\partial z} + \left(\frac{1}{v_{g3}}\frac{\partial}{\partial t} + \frac{i\beta_3}{2}\frac{\partial^2}{\partial t^2}\right)E_3 = C_3\left(E_1 E_2^* e^{i\Delta k_{123}z} + E_3^* E_6 e^{i\Delta k_{633}z} + E_1^* E_8 e^{i\Delta k_{813}z} + E_2 E_9^* e^{i\Delta k_{239}z}\right)$$
$$+ i\frac{3\omega_3 \chi^{(3)}}{8n_3 c} E_3\left(S - |E_3|^2\right)$$

$$\frac{\partial E_4}{\partial z} + \left(\frac{1}{v_{g4}}\frac{\partial}{\partial t} + \frac{i\beta_4}{2}\frac{\partial^2}{\partial t^2}\right)E_4 = \frac{1}{2}C_4 E_1^2 e^{-i\Delta k_{411}z} + i\frac{3\omega_4 \chi^{(3)}}{8n_4 c} E_4\left(S - |E_4|^2\right)$$

$$\frac{\partial E_5}{\partial z} + \left(\frac{1}{v_{g5}}\frac{\partial}{\partial t} + \frac{i\beta_5}{2}\frac{\partial^2}{\partial t^2}\right)E_5 = \frac{1}{2}C_5 E_2^2 e^{-i\Delta k_{522}z} + i\frac{3\omega_5 \chi^{(3)}}{8n_5 c} E_5\left(S - |E_5|^2\right)$$

$$\frac{\partial E_6}{\partial z} + \left(\frac{1}{v_{g6}}\frac{\partial}{\partial t} + \frac{i\beta_6}{2}\frac{\partial^2}{\partial t^2}\right)E_6 = \frac{1}{2}C_6 E_3^2 e^{-i\Delta k_{633}z} + i\frac{3\omega_6 \chi^{(3)}}{8n_6 c} E_6\left(S - |E_6|^2\right)$$

$$\frac{\partial E_7}{\partial z} + \left(\frac{1}{v_{g7}}\frac{\partial}{\partial t} + \frac{i\beta_7}{2}\frac{\partial^2}{\partial t^2}\right)E_7 = C_7 E_1 E_2 e^{-i\Delta k_{712}z} + i\frac{3\omega_7 \chi^{(3)}}{8n_7 c} E_7\left(S - |E_7|^2\right)$$

$$\frac{\partial E_8}{\partial z} + \left(\frac{1}{v_{g8}}\frac{\partial}{\partial t} + \frac{i\beta_8}{2}\frac{\partial^2}{\partial t^2}\right)E_8 = C_8 E_1 E_3 e^{-i\Delta k_{813}z} + i\frac{3\omega_8 \chi^{(3)}}{8n_8 c} E_8\left(S - |E_8|^2\right)$$

$$\frac{\partial E_9}{\partial z} + \left(\frac{1}{v_{g9}}\frac{\partial}{\partial t} + \frac{i\beta_9}{2}\frac{\partial^2}{\partial t^2}\right)E_9 = C_9 E_2 E_3^* e^{i\Delta k_{239}z} + i\frac{3\omega_9 \chi^{(3)}}{8n_9 c} E_9\left(S - |E_9|^2\right)$$

where $S = 2\sum_{j=1}^{9}|E_k|^2 = 2\left(|E_1|^2 + |E_2|^2 + \cdots + |E_9|^2\right)$, $\chi^{(3)}$ is the third-order nonlinear susceptibility of LiNbO$_3$. Since the value of $\chi^{(3)}$ is far smaller than that of $\chi^{(2)}$ in LiNbO$_3$, unless the light intensity is very high, the $\chi^{(3)}$ effects, such as self-phase modulation (SPM) and cross-phase modulation (XPM), are negligible in the $\chi^{(2)}$ crystal. The simulation shows that the addition of $\chi^{(3)}$ terms will increase the computation time by 40%. For these reasons, we only discuss second-order nonlinearities in this work.

## 6. Experimental support for nine-wave coupled equations

There is an ongoing experiment of OPO in our lab that supports the prediction of the nine-wave coupled equations. In this experiment, the OPO is pumped by a Yb-doped femtosecond laser with 1030nm central wavelength, 120fs pulse width and 3.1GW/cm$^2$ peak intensity (focused). The nonlinear crystal is a 1mm PPLN, which poling period is for the main OPO process (1030nm→1300nm+4959nm). The transformation relationship of the nine waves is shown in Figure S3.

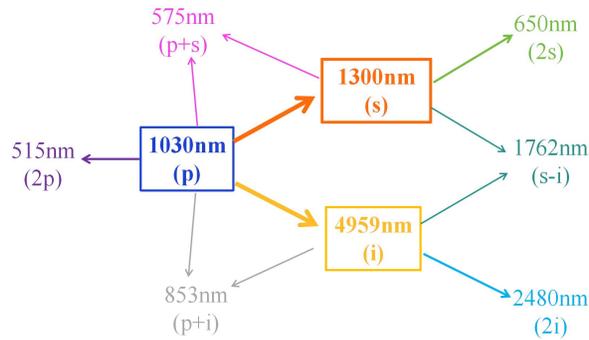

Figure S3. Schematic diagram of the nine-wave interaction relationship for 1300nm signal.

The evolution of the optical field in OPO is simulated using the nine-wave coupled equations. The tepmoral and spectral domain results of the output optical fields are shown in Figure S4(a) and (b). It can be seen that, besides the pump, signal, and idler, a strong optical field near 575nm is generated, which is the sum-frequency of the pump and the signal. This phenomenon cannot be reveal by traditional three-wave coupled equations. In the 4 strong output fields (1030nm, 1300nm, 4959nm and 575nm), only the 575nm (yellow color light) is in the visible regime. This yellow light is predicted by the nine-wave coupled equantions, and has been confirmed by the experiment. (Figure S4(c)).

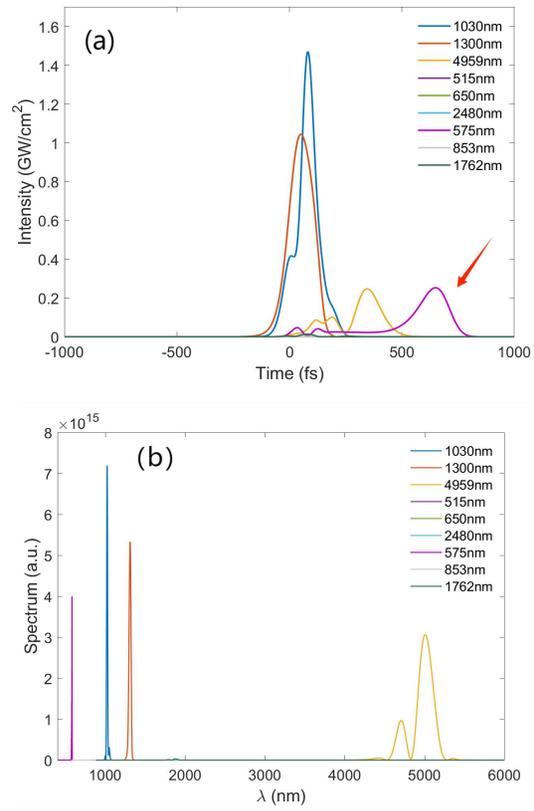

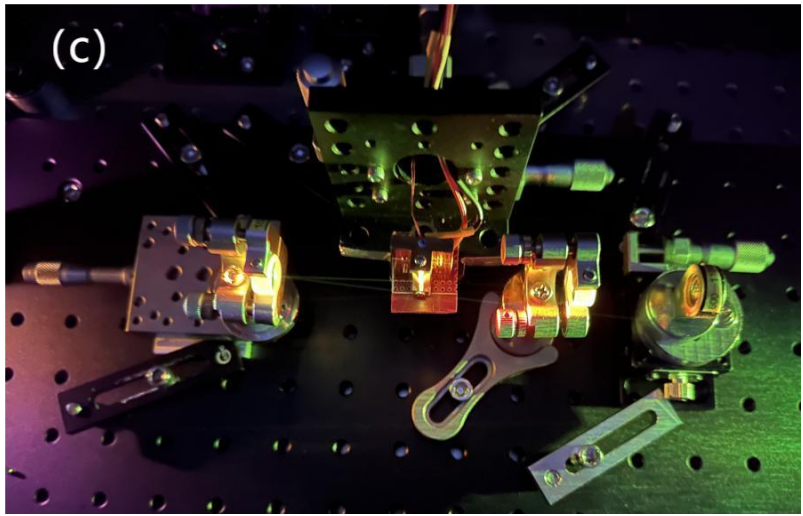

Figure S4. Temporal (a) and spctral (b) domain results of a 1300nm OPO simulated by nine-wave coupled equations. (c) Experiment photo of the OPO.